\documentstyle[aps,prl,twocolumn,epsf,floats]{revtex}
\begin{document}
\draft
\wideabs{
\title{Excitonic Ions and Pseudopotentials in 2D Systems:\protect\\
       Evidence for Quantum Hall States of an $X^-$ Gas}
\author{
   Arkadiusz W\'ojs$^{a*}$, Pawel Hawrylak$^b$, and John J. Quinn$^{a*}$}
\address{
   $^a$Department of Theoretical Physics, 
       University of Tennessee, Knoxville, Tennessee 37996-1501 \\
   $^b$Institute for Microstructural Sciences, 
       National Research Council of Canada, Ottawa, Canada K1A 0R6}

\maketitle

\begin{abstract}
   Systems of up to twelve electrons and six holes on the Haldane 
   sphere are studied by exact numerical diagonalization.
   The low lying states of the system involve bound excitonic 
   complexes such as $(X^n)^-$.
   The angular momenta of these complexes and the pseudopotentials 
   describing their interaction are determined.
   The similarity to the electron pseudopotential suggests the 
   possibility of incompressible ground states of a gas of $X^-$
   ions for $\nu\le1/3$.
   The $\nu=1/3$ state of three $X^-$'s is found.
\end{abstract}
\pacs{71.10.Pm, 73.20.Dx, 73.40.Hm, 71.35.Ji}

}

\paragraph*{Introduction.}
In a quasi-two-dimensional system in the presence of a dc magnetic 
field a pair of electrons ($e^-$) and a valence band hole ($h^+$) 
can form a negatively charged exciton ($X^-$) 
\cite{kheng,buhmann,chen,wojs1,palacios}.
This state has lower energy than the multiplicative state predicted 
by hidden symmetry \cite{lerner}, which consists of a neutral exciton 
($X$) in its ground state and an unbound free electron.
Generally, in the system of electrons and holes confined to their lowest 
(spin polarized) Landau levels, the only bound complexes are $X$ and 
charged multi-exciton complexes, or excitonic ions, $(X^n)^-$ ($n=1$, 
2, \dots).
Moreover, each $(X^n)^-$ ion has only one bound state and the binding 
energy, $\Delta_{(X^n)^-}=E_{(X^{n-1})^-}+E_X-E_{(X^n)^-}$, quickly 
decreases with increasing $n$.
The $X^-$ and larger ions $(X^n)^-$ are long-lived \cite{palacios} 
composite particles with mass and charge; therefore their lowest energy 
states form a degenerate Landau level \cite{wojs1}.
It seems plausible that such composite particles could form Laughlin
\cite{laughlin} incompressible ground states at particular values of 
magnetic field \cite{wojs2}.

In this paper we study by exact numerical diagonalization the energy 
spectra of small systems containing $N_h$ holes and $N_e$ electrons
($N_e>N_h$), confined to the surface of a Haldane sphere \cite{haldane} 
of radius $R$ and monopole strength $2S(hc/e)$.
From our numerical results we are able to determine the angular momentum 
$l_A$ of the composite particles ($A=X^-$, $(X^2)^-$, etc.) and the 
pseudopotentials $V_{AB}(L)$ describing the interaction of any pair 
$AB$ as a function of the pair angular momentum $L$.
At sufficiently large values of $S$ and small values of $L$, the 
pseudopotentials of all pairs are very similar and can be well 
approximated by those of a pair of electrons (point particles) with 
individual angular momenta $l_A$ and $l_B$.
However, if $A$ or $B$ is a composite particle, the maximum allowed pair 
angular momentum is smaller than that of two point particles with angular
momenta $l_A$ and $l_B$.
This is equivalent to a ``hard core'' repulsion, effectively raising 
one or more of the highest pseudopotential parameters to infinity.

Knowing binding energies and angular momenta of the composite particles 
and the $V_{X^-X^-}(L)$ pseudopotential allows us to use the composite 
Fermion (CF) picture \cite{jain} to predict the lowest lying band of 
angular momentum multiplets of a system of $X^-$ particles for various 
values of the magnetic field.
These predictions are compared with exact numerical results for a system
of six electrons and three holes at $2S\le11$ (three $X^-$ particles
at $\nu_{X^-}>1/5$; filling factor $\nu_{X^-}$ will be defined later), 
and the agreement is found to be good.
For larger systems it becomes difficult to carry out exact numerical
diagonalization in terms of the individual electrons and holes, but the
lowest lying bands of states will consist of $X^-$'s interacting through 
the pseudopotential $V_{X^-X^-}(L)$, which can be handled numerically.

\paragraph*{Model.}
The single particle states on the Haldane sphere are called monopole 
harmonics \cite{wuyang}.
They are labeled by angular momentum $l\ge S$ and its projection $m$.
The lowest Landau level consists of the $l=S$ multiplet.
The many particle Hilbert space is spanned by single particle 
configurations classified by the total angular momentum projection $M$.
The many particle eigenstates, obtained through numerical diagonalization, 
fall into degenerate total angular momentum ($L$) multiplets.

\begin{figure}[t]
\epsfxsize=3.35in
\epsffile{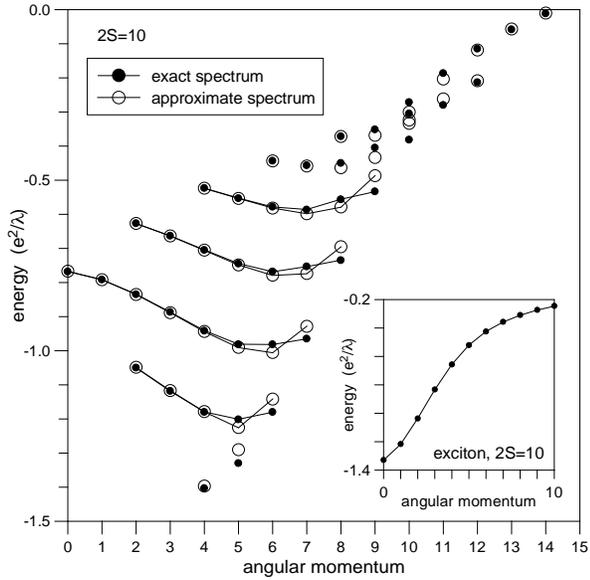}
\caption{
   Energy spectrum of two electrons and one hole at $2S=10$.
   Solid circles -- exact spectrum;
   open circles -- approximate spectrum.
   Inset: energy spectrum of an electron--hole pair.}
\label{fig1}
\end{figure}

\paragraph*{Two Electron--One Hole System.}
In Fig.~\ref{fig1} we show as solid circles the energy spectrum of 
a system of two electrons and one hole at $2S=10$ as a function of the 
total angular momentum $L$.
The lowest energy state at $L=S$ is the multiplicative state with one 
exciton in its $l_X=0$ ground state and one electron of angular momentum 
$l_e=S$.
Only one state of lower energy occurs in the spectrum.
It appears at $L=S-1$ and corresponds to the only bound state of the 
negatively charged exciton $X^-$ \cite{wojs2}.
The value of the $X^-$ angular momentum, $l_{X^-}=S-1$, can be understood
by noticing that the lowest energy single particle configuration of two 
electrons and one hole is the ``compact droplet'', in which the two 
electrons have $m=S$ and $m=S-1$, and the hole has $m=-S$, giving $M=S-1$.

As marked with lines in Fig.~\ref{fig1}, unbound states above the 
multiplicative state form bands, which arise from the $e$--$h$ 
interaction and are separated by gaps associated with the characteristic 
excitation energies of an $e$--$h$ pair (the $e$--$h$ pseudopotential, 
i.e. the energy spectrum of an exciton, is shown in the inset).
These bands are well approximated (open circles) by the expectation 
values of the total ($e$--$e$ and $e$--$h$) interaction energy, 
calculated in the eigenstates of the $e$--$h$ interaction alone.

\begin{figure}[t]
\epsfxsize=3.35in
\epsffile{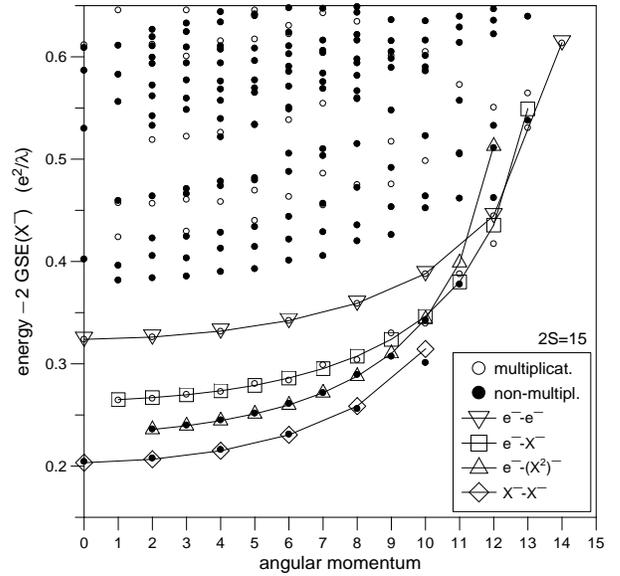}
\caption{
   Energy spectrum of four electrons and two holes at $2S=15$.
   Open circles -- multiplicative states;
   solid circles  -- non-multiplicative states;
   triangles, squares, and diamonds -- approximate pseudopotentials.}
\label{fig2}
\end{figure}

\paragraph*{Four Electron--Two Hole System.}
In Fig.~\ref{fig2} we display the energy spectrum of a system of four
electrons and two holes at $2S=15$.
The states marked by open and solid circles are multiplicative (containing
one or more decoupled $X$'s) and non-multiplicative states, respectively.
For $L<10$ there are four rather well defined low-lying bands.
Two of them begin at $L=0$.
The lower of these consists of two $X^-$ ions interacting through
a pseudopotential $V_{X^-X^-}(L)$.
The upper band consists of states containing two decoupled $X$'s
plus two electrons interacting through $V_{e^-e^-}(L)$.
The band that begins at $L=1$ consists of one $X$ plus an $X^-$
and an electron interacting through $V_{e^-X^-}(L)$, while
the band which starts at $L=2$ consists of an $(X^2)^-$ interacting 
with a free electron.

Remember that $l_e=S$, $l_{X^-}=S-1$, and $l_{(X^2)^-}=S-2$, and that
decoupled excitons do not carry angular momentum ($l_X=0$).
For a pair of identical Fermions of angular momentum $l$ the allowed
values of the pair angular momentum are $L=2l-j$, where $j$ is an
odd integer.
For a pair of distinguishable particles with angular momenta $l_A$ and 
$l_B$, the total angular momentum satisfies $|l_A-l_B|\le L\le l_A+l_B$.
The states containing two free electrons and two decoupled neutral 
excitons fit exactly the pseudopotential for a pair of electrons at
$2S=15$; the maximum pair angular momentum is $L^{\rm MAX}=14$ as expected.
The states containing two $X^-$'s terminate at $L=10$.
Since the $X^-$'s are Fermions, one would have expected a state at 
$L^{\rm MAX}=2l_{X^-}-1=12$.
This state is missing in Fig.~\ref{fig2}.
By studying $2X^-$ states for low values of $S$, we surmise that the 
state with $L=L^{\rm MAX}$ does not occur because of the finite size 
of the $X^-$.
Large pair angular momentum corresponds to the small average distance,
and two $X^-$'s in the state with $L^{\rm MAX}$ would be too close to 
one another for the bound $X^-$'s to remain stable.
We can think of this as a ``hard core'' repulsion for $L=L^{\rm MAX}$.
Effectively, the corresponding pseudopotential parameter, $V_{X^-X^-}
(L^{\rm MAX})$ is infinite.
In a similar way, $V_{e^-X^-}(L)$ is effectively infinite for 
$L=L^{\rm MAX}=14$, and $V_{e^-(X^2)^-}(L)$ is infinite for 
$L=L^{\rm MAX}=13$.

Once the maximum allowed angular momenta for all four pairings $AB$ are 
established, all four bands in Fig.~\ref{fig2} can be well approximated 
by the pseudopotentials of a pair of electrons (point charges) with 
angular momenta $l_A$ and $l_B$, shifted by energies of appropriate 
composite particles.
For example, the $X^-$--$X^-$ band is approximated by the $e^-$--$e^-$
pseudopotential for $l=l_{X^-}=S-1$ plus twice the $X^-$ energy, and
the $e^-$--$X^-$ band is approximated by the $e^-$--$e^-$ pseudopotential 
for one electron with $l=l_e=S$ and the other with $l=l_{X^-}=S-1$, 
plus the energies of $X^-$ and of the decoupled $X$.

The $e^-$--$e^-$ pseudopotentials used to model the interaction between 
pairs of particles of angular momenta $l_A$ and $l_B$ must be calculated 
for an effective value of $2S$ given by $2S'=l_A+l_B$.
The model pseudopotential (obtained in units of $e^2/\lambda(S')$, where
$\lambda(S')=R/\sqrt{S'}$ is the magnetic length seen by electrons) must
be taken in units of $e^2/\lambda(S)$ to describe $V_{AB}$.
The agreement is demonstrated in Fig.~\ref{fig2}, where the squares, 
diamonds, and two kinds of triangles approximate the four bands in 
the four electron--two hole spectrum.
The fit of the diamonds to the actual $X^-$--$X^-$ spectrum is quite
good for $L<10$.
The fit of the $e^-$--$X^-$ squares to the open circle multiplicative 
states is reasonably good for $L<12$, and the $e^-$--$(X^2)^-$ triangles
fit their solid circle non-multiplicative states rather well for $L<11$.

At sufficiently large separation (low $L$), the repulsion between ions 
is weaker than their binding and the ion--ion scattering does not excite 
their internal degrees of freedom.
Consequently, the low lying states can be viewed as pure ion--ion 
excitations, and coupling to the internal dynamics of ions requires 
higher energies.

\paragraph*{Many $X^-$ System.}
We know from exact diagonalization calculations for up to ten electrons 
\cite{fano} that the CF picture correctly predicts the low lying states 
of the fractional quantum Hall systems.
The reason for this success has been shown \cite{wojs3} to be the ability
of the electrons in states of low total angular momentum to avoid large 
fractional parentage from pair states associated with large repulsive 
values of the Coulomb pseudopotential $V_{e^-e^-}(L)$.
In particular, for the Laughlin $\nu=1/3$ (or 1/5) state, the fractional
parentage from pair states with the pair angular momentum of $L^{\rm MAX}$ 
(or both $L^{\rm MAX}$ and $L^{\rm MAX}$-2) vanishes.
We hypothesize that the same effect should occur for a system of $X^-$'s
when $l_e=S$ is replaced by $l_{X^-}=S-1$.
Defining the effective $X^-$ filling factor of the $NX^-$ system at the
monopole strength of $2S$ as $\nu_{X^-}(N,S)=\nu(N,S-1)$, where $\nu$ is 
the electron filling factor, we expect the occurrence of incompressible 
$L=0$ ground states of negatively charged excitons at all Laughlin and 
Jain fractions for $\nu_{X^-}\le1/3$.
States with $\nu_{X^-}>1/3$ cannot be constructed because they have some 
fractional parentage from pair angular momentum $L^{\rm MAX}$, which is 
essentially infinite due to the hard core repulsion.

\begin{figure}[t]
\epsfxsize=3.35in
\epsffile{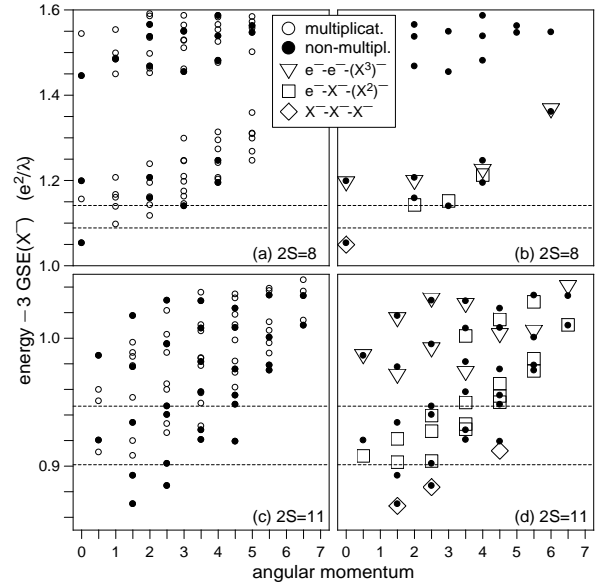}
\caption{
   Energy spectra of six electrons and three holes at $2S=8$ and 11.
   Open circles -- multiplicative states;
   solid circles -- non-multiplicative states;
   triangles, squares, and diamonds -- approximate spectra;
   dashed lines -- estimated lower bounds of spectra given by
   triangles and squares.}
\label{fig3}
\end{figure}

In Fig.~\ref{fig3} we present numerical results for a system of six 
electrons and three holes, for values of $2S=8$ and $2S=11$.
Both multiplicative (open circles) and non-multiplicative (solid circles)
states are shown in frames (a) and (c).
In frames (b) and (d) only the non-multiplicative states are plotted,
together with the approximate spectra marked with large open symbols.
The approximate spectra are obtained by diagonalizing the system of three
electrons with angular momenta appropriate for three possible groupings 
of six electrons and three holes into three ions:
$X^-$--$X^-$--$X^-$ (diamonds), $e^-$--$X^-$--$(X^2)^-$ (squares), and 
$e^-$--$e^-$--$(X^3)^-$ (triangles), and using the model pseudopotentials 
discussed earlier.

Good agreement between the exact and approximate spectra in 
Figs.~\ref{fig3}b and \ref{fig3}d allows identification of the three 
ion states in the spectrum.
States corresponding to different groupings form bands.
The bands overlap slightly at higher $L$, but at low $L$ they are 
separated by gaps reflecting different energies of ions in different 
groupings.
The lowest energy state within each band corresponds to the three ions
moving as far from each other as possible (maximally avoiding high pair 
angular momenta).
If the ion--ion repulsion energies were equal for all groupings (a good 
approximation for dilute systems), the two higher bands would lie above 
dashed lines, marking the ground state energy plus the appropriate 
difference in binding energies.
The low lying multiplicative states in Figs.~\ref{fig3}a and \ref{fig3}c 
can also be identified as three ion states and fall into following bands: 
$3X$--$e^-$--$e^-$--$e^-$, $2X$--$e^-$--$e^-$--$X^-$, 
$X$--$e^-$--$e^-$--$(X^2)^-$, and $X$--$e^-$--$X^-$--$X^-$,
whose bottom edges can be estimated based on binding energies.
The bands of three ion states are separated by a rather large gap from all 
other states, which involve excitation and breakup of composite particles.

It follows from above discussion that the energy spectrum of $N_e$ 
electrons and $N_h$ holes contains well developed low energy bands 
of states containing particular combinations of bound charged composite 
particles (ions) and decoupled excitons.
The relative position of bands is governed by the binding energies 
of ions.
The $(X^n)^-$ binding energy decreases sufficiently quickly with 
increasing $n$, that if $nN_e=(n+1)N_h$, the lowest band consists 
of states of $N_e-N_h$ identical $(X^n)^-$ ions.

Knowing that the lowest lying states in Fig.~\ref{fig3} contain three 
$X^-$'s (states approximated by the diamonds) and using the arguments 
on the correspondence between the $Ne^-$ system at the monopole strength 
$2S$ and the $NX^-$ system at $2(S-1)$, we can make the following 
identification:
(i)
The lowest energy state at $2S=8$ is an $L=0$ incompressible ground state 
of three $X^-$'s corresponding to $\nu_{X^-}=1/3$.
In fact this is the only state of three $X^-$'s for this value of $2S$.
Other states of three $X^-$'s would involve some fractional parentage 
from pair angular momentum $L^{\rm MAX}$, which is forbidden by the 
hard core repulsion.
(ii)
The lowest energy state at $2S=11$ corresponds to one quasielectron 
with $l_{QE}=3/2$ in the $\nu=1/5$ Laughlin state, and thus can be 
thought of as one quasi-$X^-$ in the $\nu_{X^-}=1/5$ state.

\begin{figure}[t]
\epsfxsize=3.35in
\epsffile{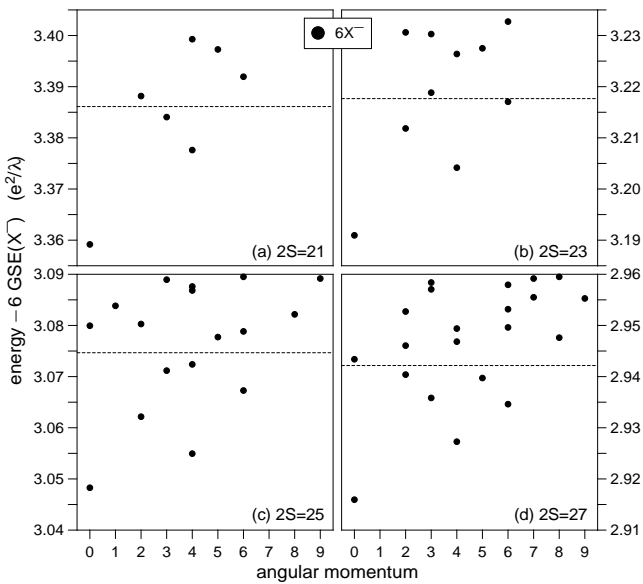}
\caption{
   Approximate $6X^-$ bands in the spectra of twelve electrons and 
   six holes at $2S=21$, 23, 25, and 27.
   Dashed lines -- estimated lower bounds of higher bands.}
\label{fig4}
\end{figure}

For larger systems it becomes quite difficult to perform exact
diagonalization at values of $2S$ corresponding to $\nu_{X^-}\le1/3$.
For example, for the twelve electron--six hole system we expect the 
$\nu_{X^-}=1/3$, 2/7, 2/9, and 1/5 incompressible states to occur at 
$2S=17$, 21, 23, and 27, respectively.
At $2S=25$ we expect four low energy states at $L=0$, 2, 4, and 6, 
which would describe two quasi-$X^-$'s, each with $l_{QX^-}=7/2$, 
in the $\nu_{X^-}=1/5$ state.
Unfortunately, the exact diagonalization of this eighteen particle system
at such values of $2S$ is beyond our current computer capability.
However, we managed to extrapolate the $V_{X^-X^-}(L)$ pseudopotential
making use of its very regular dependence on $2S$, and use it to determine 
approximate bands of $6X^-$ states.
The spectra obtained in this way are shown in Fig.~\ref{fig4}.
We know that the $6X^-$ band is the lowest energy band in the twelve 
electron--six hole spectrum.
We expect that the next lowest band is $e^-$--$4X^-$--$(X^2)^-$ and that 
it begins about the energies given by the dashed lines, equal to the 
ground state energies plus $E_{(X^2)^-}-2E_{X^-}$.
All features predicted by the CF picture for six electrons can be seen 
in Fig.~\ref{fig4}.

Note also that the number of $NX^-$ states lying below higher bands,
corresponding to other combinations of ions, increases when $\nu_{X^-}$ 
decreases.
This is because the ion binding energies increase and the ion--ion 
repulsion energy decreases when $\nu_{X^-}$ decreases.
At sufficiently low $\nu_{X^-}$, a significant low energy part of the 
spectrum remains unaffected by the possibility of appearance of other 
combination of ions.
As in an electron system, the excitation gaps above the incompressible 
ground states are related to the pseudopotential and not to the ion 
binding energies.
Hence, these gaps are not expected to collapse in the thermodynamic 
limit and the $X^-$ gas is predicted to exhibit the fractional quantum 
Hall effect.

\paragraph*{Conclusion.}
We have demonstrated that the low lying states of a system of electrons 
and holes involve bound composite particles (neutral excitons, charged 
excitons, and charged multi-excitons) and free electrons.
We have obtained the pseudopotentials describing the interactions of
pairs of charged composite particles and used it to study the low lying
states of three and six $X^-$'s.
Laughlin incompressible states are found for $\nu_{X^-}\le1/3$.
Results for the $3X^-$ system agree well with exact numerical results
for six electrons and three holes.

\paragraph*{Acknowledgment.}
We thank J. J. Palacios (Universidad Autonoma de Madrid) and 
M. Potemski (Grenoble High Magnetic Field Laboratory) for helpful
discussions.

\end{document}